\documentclass[reqno,11pt,a4paper]{amsart}

\let\phi=\varphi
\DeclareMathOperator{\rank}{rank}
\newcommand*{\sd}[2]{\{\,#1\mid#2\,\}}
\newcommand*{\eval}[1]{\left.#1\right|}
\newcommand*{\abs}[1]{\left|#1\right|}
\newcommand*{\Ev}{\mathbf{E}}
\theoremstyle{theorem}
\newtheorem{proposition}{Proposition}
\theoremstyle{definition}
\newtheorem*{coord}{Coordinates}
\newtheorem*{example}{Example}
\usepackage[all]{xy}
\SelectTips{cm}{}

\usepackage{mathrsfs}
\let\mathcal\mathscr
\usepackage[mathscr]{eucal}

\newcommand{\cprime}{\/{\mathsurround=0pt$'$}}

\begin{document}

\title[A natural geometric construction]{A natural geometric construction
    underlying a class of Lax pairs} \author{I.S.~Krasil{\cprime}shchik}
  \address{Slezsk\'{a} univerzita v Opav\v{e}, Matematick\'{y} \'{u}stav v
    Opav\v{e}, Na Rybn\'{\i}\v{c}ku 626/1, 746 01 Opava, Czech Republic \&
    Independent University of Moscow, 119002, B.~Vlasyevskiy Per. 11, Moscow,
    Russia}\thanks{I am grateful to the Mathematical Institute of the Silesian
    University in Opava for support and comfortable working condition.}
  \email{josephkra@gmail.com}


\begin{abstract}
  In the framework of the theory of differential coverings~\cite{KV}, we
  discuss a general geometric construction that serves the base for the
  so-called Lax pairs containing differentiation with respect to the spectral
  parameter~\cite{OS}. Such kind of objects arise, for example, when studying
  integrability properties of equations like the Gibbons-Tsarev one~\cite{GT}.
\end{abstract}



\maketitle

\section*{Introduction}
The \emph{Gibbons-Tsarev equation}
\begin{equation}
z_{yy}+z_xz_{xy}-z_yz_{xx}+1=0,\label{eq:10}
\end{equation}
see, e.g.,~\cite{GT}, arises as a reduction of the \emph{Benney chain} and
possesses many properties of integrable systems. On the other hand, it is not
integrable in the exact solitonic sense (for example, the equation does not
admit Hamiltonian operators). In particular, equation~\eqref{eq:10} is the
compatibility condition for the system
\begin{equation}\label{eq:1}
  \phi_x=\frac{1}{z_y+z_x\phi-\phi^2},\quad
  \phi_y=-\frac{z_x-\phi}{z_y+z_x\phi-\phi^2}.
\end{equation}
Introducing a new function~$\psi=\psi(\phi)$, one can consider the system (`a
Lax pair')
\begin{equation}\label{eq:2}
  \psi_x=-\frac{1}{z_y+z_x\phi-\phi^2}\psi_\phi,\quad
  \psi_y=\frac{z_x-\phi}{z_y+z_x\phi-\phi^2}\psi_\phi
\end{equation}
(see, for example,~\cite{OS}), whose compatibility condition is the
Gibbons-Tsarev equation as well. Taking now~$\phi$ for a parameter
in~\eqref{eq:2} and expanding the resulting system in Laurent series, one can
describe an infinite family of nonlocal conservation laws and the
corresponding infinite algebra of nonlocal symmetries.

Of course, these computations can be done in a more general
situation. Let~$\mathcal{E}$ be a PDE imposed on an unknown function~$z(x,y)$
and~$\phi_x=X(x,y,[z],\phi)$, $\phi_y=Y(x,y,[z],\phi)$ be a system compatible
by virtue of~$\mathcal{E}$, where~$[z]$ denotes a collection consisting of~$z$
itself and its derivatives up to a certain order. Then one can pass to the
system~$\psi_x=-X(x,y,[z],\phi)\psi_\phi$, $\psi_y=-Y(x,y,[z],\phi)\psi_\phi$
also compatible over~$\mathcal{E}$. As I see it myself, this seemingly simple
construction invokes two questions at least: (a)~why do the signs change?
(b)~how does the unknown function~$\phi$ in the initial system transform to an
independent variable in the resulting one?

Below we shall discuss a general geometric construction that (hopefully)
answers these questions and explains the trick of passing from
System~\eqref{eq:1} to System~\eqref{eq:2}. In Section~\ref{sec:-main} we
briefly recall the necessary definitions and results from the geometric theory
of differential equations (see~\cite{VKL}, for example), including the
nonlocal theory,~\cite{KV}. Section~\ref{sec:-jets-dist} concerns with the
definition of jet spaces associated to integrable distributions. Finally, in
Section~\ref{sec:-jets-cov} the main construction is described.

\section{The basic notions and notation}
\label{sec:-main}

Consider a smooth manifold~$M$, $\dim M=n$, and a locally trivial vector fiber
bundle~$\pi\colon E\to M$, $\rank\pi=m$. Denote by~$\Gamma(\pi)$ the
$C^\infty(M)$-module of sections of the bundle~$\pi$. Let~$\pi_k\colon
J^k(\pi)\to M$ be the corresponding bundle of $k$-jets, $k=1,\dots,\infty$,
and let~$\pi_{k,l}\colon J^k(\pi)\to J^l(\pi)$, $k>l$, denote the natural
projections. The bundle~$\pi_\infty$ possesses a natural flat
connection~$\mathcal{C}$, which is called the \emph{Cartan connection} and is
defined by the condition
\begin{equation*}
  j_\infty(f)^*(\mathcal{C}_X(\phi))=X(j_\infty(f)^*(\phi)),
\end{equation*}
where~$X$ is a vector field on~$M$, $\phi$ is a smooth function
on~$J^\infty(\pi)$, $f$ is a section of~$\pi$,
and~$j_\infty(f)\in\Gamma(\pi_\infty)$ denotes the infinite jet of this
section. The corresponding horizontal distribution~$z\mapsto\mathcal{C}_z$,
$z\in J^\infty(\pi)$, on~$J^\infty(\pi)$ is integrable and is called the
\emph{Cartan distribution}.

An (\emph{infinitely prolonged}) \emph{differential equation} will be
understood as a submanifold~$\mathcal{E}\subset J^\infty(\pi)$ such that all
the fields of the form~$\mathcal{C}_X$ are tangent to this
submanifold. Consequently, the Cartan connection can be restricted to the
projection~$\eval{\pi_\infty}_{\mathcal{E}}$, while the Cartan distribution is
restricted to~$\mathcal{E}$. We shall keep the same notation for these
restrictions.

Let~$\tilde{\mathcal{E}}\subset J^\infty(\tilde{\pi})$,
$\tilde{\pi}\colon\tilde{E}\to M$, be another equation. A smooth vector
bundle~$\tau\colon\tilde{\mathcal{E}}\to\mathcal{E}$ is called a
(\emph{differential}) covering if one
has~$\tau_*\circ\tilde{\mathcal{C}}_X=\mathcal{C}_X$ for any vector
field~$X\in D(M)$, where~$D(M)$ denotes the module of vector fields on~$M$ and
`star' stands for the differential of a smooth map.

\begin{coord}
  Let~$\mathcal{U}\subset M$ be a local chart in~$M$ with
  coordinates~$x^1,\dots,x^n$
  and~$\pi^{-1}(\mathcal{U})=\mathcal{U}\times\mathbb{R}^m$ be a
  trivialization of the bundle~$\pi$ with some coordinates~$u^1,\dots,u^m$ in
  the fibers. Then the \emph{adapted coordinates}~$u_\sigma^j$ arise
  in~$J^\infty(\pi)$. The Cartan connection in these coordinates is defined by
  the \emph{total derivative} operators
  \begin{equation*}
    D_{x^i}=\frac{\partial}{\partial x^i} + \sum_{j,\sigma}u_{\sigma
      i}^j\frac{\partial}{\partial u_\sigma^j},
  \end{equation*}
  where~$\sigma$ is the multi-index corresponding to the
  variables~$x^i$. A covering structure in the
  bundle~$\tau\colon\tilde{\mathcal{E}}\to\mathcal{E}$ with
  coordinates~$w^1,\dots,w^r$ along the fibers (the functions~$w^\gamma$ are
  called \emph{nonlocal variables}) is given by the extensions of the total
  derivatives
  \begin{equation}\label{eq:4}
    \tilde{D}_{x^i}=D_{x^i}+X_i,
  \end{equation}
  where~$X_i$ are $\tau$-vertical vector fields (i.e., the fields of the
  form~$\sum_\gamma X_i^\gamma\partial/\partial w^\gamma$) that satisfy the
  relations
  \begin{equation*}
    D_{x^i}(X_j)-D_{x^j}(X_i)+[X_i,X_j]=0
  \end{equation*}
  for all~$i<j$.
\end{coord}

A covering~$\tau$ is called \emph{trivial}
if~$\mathcal{C}_X(C^\infty(\tilde{\mathcal{E}}))\subset C^\infty(\mathcal{E})$
for all vector fields~$X\in D(M)$. A \emph{morphism} of two
coverings~$\tau_i\colon\tilde{\mathcal{E}}_\alpha\to\mathcal{E}$, $\alpha=1$,
$2$, of ranks~$r_1$ and~$r_2$, respectively, is a smooth
map~$\phi\colon\tilde{\mathcal{E}}_1\to\tilde{\mathcal{E}}_2$ such that
\begin{enumerate}
\item $\tau_1=\tau_2\circ\phi$ and
\item $\phi_*(\tilde{\mathcal{C}}_z^1)\subset\tilde{\mathcal{C}}_{\phi(z)}^2$,
  $z\in\tilde{\mathcal{E}}$.
\end{enumerate}

\begin{coord}
  A covering~$\tau$ is trivial if and only if the vertical fields in
  Equalities~\eqref{eq:4} vanish.

  If~$w_\alpha^j$ are the nonlocal variables in the covering~$\tau_\alpha$,
  $\alpha=1$, $2$, while the covering structures are given by the vertical
  fields
  \begin{equation*}
    \sum_\gamma X_i^{\alpha,\gamma}\frac{\partial}{\partial w_\alpha^\gamma}
  \end{equation*}
  then a morphism~$\phi$ of~$\tau_1$ to~$\tau_2$ is determined by a system of
  functions~$w_1^\gamma=\phi^\gamma(w_1^1,\dots,w_1^{r_1})$,
  $\gamma=1,\dots,r_2$, such
  that~$\tilde{D}_{x^i}^1(\phi^\gamma)=\phi^*(X_i^{2,\gamma})$. In
  particular,~$\phi$ is a morphism to the trivial covering if
  \begin{equation}
    \label{eq:6}
    \tilde{D}_{x^i}(\phi^\gamma)=0.
  \end{equation}
  Such morphisms will be called \emph{trivializing} ones.
\end{coord}

\section{Jets associated to distributions}
\label{sec:-jets-dist}

Consider a smooth manifold~$N$, $\dim N\leq\infty$. Let~$\mathcal{D}$ be an
integrable distribution on this manifold,~$\rank\mathcal{D}=s<\infty$, i.e., a
locally projective finite-rank submodule~$\mathcal{D}\subset D(N)$ such
that~$[\mathcal{D},\mathcal{D}]\subset\mathcal{D}$. Let also~$\xi\colon F\to
N$ be a locally trivial vector bundle over~$N$. We say that two sections~$f$
and~$f'\in\Gamma(\xi)$ are \emph{$k$-equivalent} over~$\mathcal{D}$ at a
point~$z\in N$ if
\begin{equation*}
  \eval{\big(X_1\dots X_l(f)\big)}_z=\eval{\big(X_1\dots
    X_l(f')\big)}_z,\qquad
  l\leq k,
\end{equation*}
for any collection of vector
fields~$X_1,\dots,X_l\in\mathcal{D}$. Let~$[f]_z^k$ denote the equivalence
class of~$f$. The set
\begin{equation*}
  J_{\mathcal{D}}^k(\xi)=\sd{[f]_z^k}{z\in N,f\in\Gamma(\xi)},\quad
  k=0,1,\dots,\infty, 
\end{equation*}
is endowed with a natural smooth manifold structure (cf.~\cite{VKL}) and is
called the \emph{the manifold of $k$-jets associated to the
  distribution~$\mathcal{D}$}. One has the natural projections~$\xi_k\colon
J_{\mathcal{D}}^k(\xi)\to M$ and~$\xi_{k,l}\colon J_{\mathcal{D}}^k(\xi)\to
J_{\mathcal{D}}^l(\xi)$, $k>l$, and all these maps are locally trivial fiber
bundles while the functions~$j_k(f)\colon M\to J_{\mathcal{D}}^k(\xi)$,
$z\mapsto[f]_z^k$, (\emph{jets of sections}) are smooth sections of the
bundles~$\xi_k$.

\begin{example}
  If~$\mathcal{D}=D(N)$ we obtain the classical definition of jet spaces.
\end{example}

\begin{example}
  If~$N=\mathcal{E}$ is an equation and~$\mathcal{D}=\mathcal{C}$ coincides
  with the Cartan distribution we arrive to the definition of \emph{horizontal
    jets}, see~\cite{VKL}.
\end{example}

The bundle of infinite jets~$\xi_\infty$ admits a flat
\emph{$\mathcal{D}$-connection} denoted~by
\begin{equation*}
  \nabla^{\mathcal{D}}\colon\mathcal{D}\to D(J_{\mathcal{D}}^\infty(\xi))
\end{equation*}
and which is also called the \emph{Cartan connection}: for any
point~$\theta=[f]_z^\infty$, a field~$X\in\mathcal{D}$ and a function~$\phi\in
C^\infty(J_{\mathcal{D}}^\infty(\xi))$, we set
\begin{equation*}
  \eval{\nabla_X^{\mathcal{D}}(\phi)}_\theta=\eval{X(j_\infty(f)^*\phi)}_z.
\end{equation*}
The corresponding horizontal distribution on~$J_{\mathcal{D}}^\infty(\xi)$
will be denoted by~$\Delta^{\mathcal{D}}$; it is integrable,
i.e.,~$[\Delta^{\mathcal{D}},\Delta^{\mathcal{D}}]\subset\Delta^{\mathcal{D}}$.

A submanifold~$\mathcal{E}_k\subset J_{\mathcal{D}}^k(\xi)$ is called a
\emph{$\mathcal{D}$-equation} of order~$k$. As in the case of `usual'
equations, one can define the
\emph{$\mathcal{D}$-prolongations}~$\mathcal{E}_k^{(l)}\subset
J_{\mathcal{D}}^{k+l}(\xi)$ of finite and infinite orders. The
latter,~$\mathcal{E}\subset J_{\mathcal{D}}^\infty(\xi)$, will be shortly
called just a \emph{$\mathcal{D}$-equation}. The
map~$\xi_\infty\colon\mathcal{E}\to M$ inherits the Cartan connection, while
the submanifols~$\mathcal{E}$ carries the corresponding horizontal
distribution.

Let~$S$ be a $\xi_\infty$-vertical vector field on~$\mathcal{E}$; it is
called a (\emph{higher infinitesimal}) \emph{$\mathcal{D}$-symmetry} of the
equation~$\mathcal{E}$
if~$[S,\Delta^{\mathcal{D}}]\subset\Delta^{\mathcal{D}}$.

\begin{coord}
  Let~$y^1,\dots,y^k,\dots$ be local coordinates in~$N$, $v^1,\dots,v^r$ be
  coordinates if fibers of a trivialization of the bundle~$\xi$. Finally,
  let~$Y_1,\dots,Y_s$ be a local basis of the distribution~$\mathcal{D}$,
  \begin{equation*}
    [Y_i,Y_j]=\sum_k c_{ij}^kY_k,\qquad c_{ij}^k\in
    C^\infty(J_{\mathcal{D}}^\infty(\xi)). 
  \end{equation*}
  Let~$\sigma=i_1\dots i_{\abs{\sigma}}$ be a multi-index of arbitrary but
  finite length~$\abs{\sigma}$, $1\leq i_\alpha\leq s$. Let us define the
  coordinate functions~$v_\sigma^j$ on~$J_{\mathcal{D}}^\infty(\xi)$ by
  setting
  \begin{equation*}
    v_\sigma^j(\theta)=\eval{Y_{i_1}\dots Y_{i_{\abs{\sigma}}}(f^j)}_z,\qquad
    \theta=[f]_z^\infty, \quad f=(f^1,\dots,f^l).
  \end{equation*}
  Then the lifts~$\nabla_Y^{\mathcal{D}}$ of vector fields~$Y\in\mathcal{D}$
  to~$J_{\mathcal{D}}^\infty(\xi)$, i.e., the
  connection~$\nabla^{\mathcal{D}}$, are defined by the equalities
  \begin{equation*}
    \nabla_{Y_\alpha}^{\mathcal{D}}(v_\sigma^j)=v_{\alpha\sigma}^j.
  \end{equation*}
  One has the following relations
  \begin{equation}
    \label{eq:3}
    v_{\alpha\sigma}^j=v_{i_1\alpha\bar{\sigma}}^j + \sum_k c_{\alpha i_1}^k
    v_{k\bar{\sigma}}^j,
  \end{equation}
  where~$\bar{\sigma}=i_2\dots i_{\abs{\sigma}}$. The functions
  \begin{equation*}
    v_K^j=v_{\underbrace{1\dots1}_{k_1\text{times}}\,\dots\,\underbrace{s\dots
        s}_{k_s\text{times}}}.
  \end{equation*}
  may be taken for independent coordinates.
\end{coord}

Obviously, \emph{$\mathcal{D}$-coverings} of $\mathcal{D}$-equations are also
defined in a straightforward way.

\section{Jets over differential coverings}
\label{sec:-jets-cov}

Let us now consider the constructions of Section~\ref{sec:-jets-dist}
when~$N=\tilde{\mathcal{E}}$, where~$\tilde{\mathcal{E}}$ is the covering
equation in some covering~$\tau\colon\tilde{\mathcal{E}}\to\mathcal{E}$. Then
the manifold~$\tilde{\mathcal{E}}$ carries three integrable distributions:
\begin{itemize}
\item the \emph{horizontal distribution} denoted by~$\mathcal{H}$ which
  coincides with the Cartan distribution $\tilde{\mathcal{C}}$
  on~$\tilde{\mathcal{E}}$;
\item the \emph{vertical distribution}~$\mathcal{V}$ generated by
  $\tau$-vertical vector fields;
\item the \emph{total distribution}~$\mathcal{T}$ which is the sum of the
  previous two.
\end{itemize}
Respectively, one can define the jet spaces associated to these distributions
and the diagram
\begin{equation*}
  \xymatrix{
    &J_{\mathcal{H}}^\infty(\xi)\ar[dr]^{\xi_{\infty}^{\mathcal{H}}}&&\\
    J_{\mathcal{T}}^\infty(\xi)\ar[ur]^{\xi^{\mathcal{TH}}}
    \ar[dr]_{\xi^{\mathcal{TV}}}\ar[rr]^{\xi_{\infty}^{\mathcal{T}}}&&
    \tilde{\mathcal{E}}\ar[r]^\tau&\mathcal{E}\\  
    &J_{\mathcal{V}}^\infty(\xi)\ar[ur]_{\xi_{\infty}^{\mathcal{V}}}&&}
\end{equation*}
is commutative (the projections~$\xi^{\mathcal{TH}}$ and~$\xi^{\mathcal{TV}}$
are defined in an obvious way).

Note that the distribution~$\Delta^{\mathcal{T}}$, just like the original
total distribution~$\mathcal{T}$, is the sum of the
distributions~$\Delta^{\mathcal{H}}$ and~$\Delta^{\mathcal{V}}$,
where~$\Delta^{\mathcal{H}}$ and~$\Delta^{\mathcal{V}}$ are obtained by
lifting~$\mathcal{H}$ and~$\mathcal{V}$ using the Cartan connection associated
to the distribution~$\mathcal{T}$. Thus, the jet
manifold~$J_{\mathcal{T}}^\infty(\xi)$ possesses \emph{three different
  geometries}. Note also that the projection~$\xi^{\mathcal{TV}}$ is a
covering with respect to the $\mathcal{V}$-geometry while
the~$\xi^{\mathcal{TH}}$ is a covering in the $\mathcal{H}$-geometry of the
space~$J_{\mathcal{T}}^\infty(\xi)$.

Note also that for any $\mathcal{H}$-equation~$\mathcal{W}\subset
J_{\mathcal{H}}^\infty(\xi)$ one can consider its
pre-image~$(\xi^{\mathcal{TH}})^{-1}(\mathcal{W})$
in~$J_{\mathcal{T}}^\infty(\xi)$ and the infinite $\mathcal{T}$-prolongation
of the latter. The resulting $\mathcal{T}$-equation is denoted
by~$\tilde{\mathcal{W}}$ also possesses three different geometries, while the
map~$\eval{\xi^{\mathcal{TH}}}_{\tilde{\mathcal{W}}}$ is a covering in the
geometry associated to the horizontal distribution. Exactly the same situation
arises when one considers $\mathcal{V}$-equations.

\begin{coord}
  Let the covering structure in the
  bundle~$\tau\colon\tilde{\mathcal{E}}\to\mathcal{E}$ be defined by vector
  fields~\eqref{eq:4} and choose
  \begin{equation*}
    \frac{\partial}{\partial w^1},\dots,\frac{\partial}{\partial w^r}
  \end{equation*}
  for a basis in the space of $\tau$-vertical fields. Then
  \begin{equation}
    \label{eq:5}
    [\tilde{D}_{x^i},\tilde{D}_{x^j}]=0,\quad [\frac{\partial}{\partial
      w^\alpha},\frac{\partial}{\partial w^\beta}]=0,\quad
    [\frac{\partial}{\partial 
      w^\alpha},D_{x^i}]=\sum_\beta\frac{\partial X_i^\beta}{\partial
      w^\alpha}\frac{\partial}{\partial w^\beta}.
  \end{equation}
  Denote by~$\sigma$ the multi-index that corresponds to the
  variables~$x^i$. Let also the multi-index~$\rho$ correspond to the
  variables~$w^\alpha$. Then we obtain the following coordinate functions in
  the jet spaces:
  \begin{itemize}
  \item $v_\sigma^j$ in~$J_{\mathcal{H}}^\infty(\xi)$;
  \item $v_\rho^j$ in~$J_{\mathcal{V}}^\infty(\xi)$;
  \item $v_{\sigma\rho}^j$ in~$J_{\mathcal{T}}^\infty(\xi)$.
  \end{itemize}
  In these coordinates, one has
  \begin{equation*}
    \Delta_{x^i}^{\mathcal{T}}=\tilde{D}_{x^i}+\sum
    v_{i\sigma\rho}^j\frac{\partial}{\partial v_{\sigma\rho}^j},\quad
    \Delta_{w^\alpha}^{\mathcal{T}}=\frac{\partial}{\partial w^\alpha} + \sum
    v_{\alpha\sigma\rho}^j\frac{\partial}{\partial w^\alpha},
  \end{equation*}
  while the functions~$v_{i\sigma\rho}^j$, $v_{\alpha\sigma\rho}^j$ are
  computed accordingly to Equations~\eqref{eq:3} and~\eqref{eq:5}.
\end{coord}

Existence of different geometries on the
manifold~$J_{\mathcal{T}}^\infty(\xi)$ leads to different notions of a
symmetry. Below, we shall need the following one: \emph{a
  $\xi_\infty^{\mathcal{T}}$-vertical field~$S$ is called a gauge
  $\mathcal{V}$-symmetry
  if~$[S,\Delta^{\mathcal{V}}]\subset\Delta^{\mathcal{V}}$}. Similar to the
classical case, one has the following description of these symmetries:

\begin{proposition}\label{sec:prop-gauge-symm}
  Gauge $\mathcal{V}$-symmetries are in one-to-one correspondence with
  sections of the
  pull-back~$(\xi_\infty^{\mathcal{T}})^*(\xi_\infty^{\mathcal{H}})$. In local
  coordinates\textup{,} to any such a section~$h$ of the
  form~$v_\sigma^j=h_\sigma^j(\theta)$\textup{,} $\theta\in
  J_{\mathcal{T}}^\infty(\xi)$\textup{,} there corresponds the symmetry
  \begin{equation*}
    \Ev_h=
    \sum_{\rho,\gamma,j}\Delta_\rho^{\mathcal{T}}(h_\sigma^j)\frac{\partial}{\partial 
      v_{\rho\sigma}^j},
  \end{equation*}
  where~$\Delta_\rho^{\mathcal{T}}$ is the composition of the total
  derivatives~$\Delta_{w^\alpha}^{\mathcal{T}}$ corresponding to the
  multi-index~$\rho$.
\end{proposition}

If~$\mathcal{W}\subset J_{\mathcal{T}}^\infty(\xi)$ is a
$\mathcal{T}$-equation then the Lie algebra of its gauge
$\mathcal{V}$-symmetries will be denoted by~$\mathfrak{gs}(\mathcal{W})$.

\section{Jets of automorphisms}
\label{sec:-jets-auto}

Consider now a particular case of the above described constructions:
let~$\xi=\tau^*(\tau)$. Sections of this bundle are naturally identified with
morphisms of the bundle~$\tau$, i.e., with smooth
maps~$\phi\colon\tilde{\mathcal{E}}\to\tilde{\mathcal{E}}$, such
that~$\tau\circ\phi=\tau$.

Consider the submanifold
\begin{equation*}
  \overline{J_{\mathcal{H}}^\infty(\xi)}= \sd{\theta\in
    J_{\mathcal{H}}^\infty(\xi)}{\theta=[\phi]_z^\infty,\ \phi(z)=z,\
    z\in\tilde{\mathcal{E}}} 
  \subset J_{\mathcal{H}}^\infty(\xi)  
\end{equation*}
in~$J_{\mathcal{H}}^\infty(\xi)$ and the $\mathcal{H}$-equation
\begin{equation*}
  \mathcal{W}(\tau)=
  \sd{\theta\in\overline{J_{\mathcal{H}}^\infty(\xi)}}{\theta=[\phi]_z^\infty,\
  \phi\text{ is a trivializing morphism}}.
\end{equation*}
Let~$\tilde{\mathcal{W}}(\tau)\subset J_{\mathcal{T}}^\infty(\xi)$ be the
$\mathcal{T}$-prolongation of this equation endowed with the
$\mathcal{H}$-geometry, i.e., equipped with the horizontal distribution, as it
was described above. Then we obtain the infinite-dimensional covering
\begin{equation}
  \label{eq:7}
  \tilde{\tau}\colon\tilde{\mathcal{W}}(\tau)\to\mathcal{E},
\end{equation}
which is associated with the initial covering~$\tau$ in a canonical way.

\begin{coord}
  Let~$\phi_\sigma^\alpha$ be local coordinates
  in~$\overline{J_{\mathcal{H}}^\infty(\xi)}$, where~$\sigma$ is the
  multi-index corresponding to the variables~$x^1,\dots,x^n$. Then the
  equation~$\mathcal{W}(\tau)$ is determined by the relations
  \begin{equation*}
    \phi_{x^i}^\gamma\equiv\tilde{D}_{x^i}(\phi^\gamma)=0,\qquad
    i=1,\dots,\dim M,\quad\gamma=1,\dots,\rank\tau,
  \end{equation*}
  or
  \begin{equation}\label{eq:8}
    \frac{\partial\phi^\gamma}{\partial x^i} + \sum_\alpha
    X_i^\alpha\frac{\partial\phi^\gamma}{\partial w^\alpha}=0.
  \end{equation}
  The nonlocal variables in the covering~$\tilde{\tau}$ are the
  functions~$\phi_\rho^\gamma=\Delta_\rho^{\mathcal{T}}(\phi^\gamma)$,
  where~$\rho$ is the multi-index that corresponds to the
  variables~$w^1,\dots,w^r$. Then the equation~$\tilde{\mathcal{W}}(\tau)$ is
  given by the infinite system
  \begin{equation*}
    \frac{\partial\phi_\rho^\gamma}{\partial x^i} +
    \Delta_\rho^{\mathcal{T}}\Big(\sum_\alpha X_i^\alpha\phi_\alpha^\gamma\Big)=0.
  \end{equation*}
  Thus, the covering structure in the bundle~$\tilde{\tau}$ is described by
  the vector fields
  \begin{equation*}
    \Delta_{x^i}^{\mathcal{T}} = D_{x^i} +
    \sum_{\rho,\gamma}\Delta_\rho^{\mathcal{T}}\Big(\sum_\alpha
    X_i^\alpha\phi_\alpha^\gamma\Big)\frac{\partial}{\partial\phi_\rho^\gamma},
  \end{equation*}
  or
  \begin{equation}
    \label{eq:9}
    \Delta_{x^i}^{\mathcal{T}} = D_{x^i} + \Ev_{h_i},
  \end{equation}
  where
  \begin{equation*}
    h_i=\left(\sum_\alpha X_i^\alpha\phi_\alpha^1,\dots,\sum_\alpha
      X_i^\alpha\phi_\alpha^r\right), \quad i=1,\dots,n.
  \end{equation*}
\end{coord}

Summarizing the above discussion, we arrive to the following

\begin{proposition}
  The covering~$\tilde{\tau}\colon\tilde{\mathcal{W}}(\tau)\to\mathcal{E}$
  defined by equalities~\eqref{eq:9} is a zero-curvature representation with
  values in the Lie algebra of gauge
  $\mathcal{V}$-symmetries~$\mathfrak{gs}(\mathcal{W})$.
\end{proposition}

Obviously, the construction introduced here is a geometrical generalization of
the `Lax pairs' mentioned in Introduction.

\section*{Acknowledgment}
I am grateful to Dr.~Michal Marvan for discussions and fruitful cooperation.

\end{document}